\documentclass[aps,prl,groupedaddress,twocolumn,showpacs,preprintnumbers,amsmath,amssymb,floatfix]{revtex4}

\pdfoutput=1

\usepackage{graphicx}% Include figure files
\usepackage{dcolumn}% Align table columns on decimal point
\usepackage{bm}% bold math
\usepackage{color}
\usepackage{amsmath}
\usepackage{multirow}
\usepackage{natbib}

\newcommand{\Sn}{SnMo$_6$S$_8$}
\newcommand{\Pb}{PbMo$_6$S$_8$}

\begin{document}

\title{\boldmath{Multi-band Superconductivity in the Chevrel Phases SnMo$_6$S$_8$ and PbMo$_6$S$_8$}}

\author{A.P. Petrovi\'c$^1$, R. Lortz$^2$, G. Santi$^1$, C. Berthod$^1$, C. Dubois$^1$, M. Decroux$^1$, A. Demuer$^3$, A.B. Antunes$^3$, A. Par\'e$^3$, D. Salloum$^4$, P. Gougeon$^4$, M. Potel$^4$, and \O. Fischer$^1$} 
\affiliation{
$^{1}$DPMC-MaNEP, Universit\'e de Gen\`eve, Quai Ernest-Ansermet 24, 1211 Gen\`eve 4, Switzerland\\
$^{2}$Department of Physics, The Hong Kong University of Science \& Technology, Clear Water Bay, Kowloon, Hong Kong\\
$^{3}$Laboratoire des Champs Magn\'etiques Intenses CNRS, 25 rue des Martyrs, B.P. 166, 38042 Grenoble cedex 9, France \\
$^{4}$Sciences Chimiques, CSM UMR CNRS 6226, Universit\'e de Rennes 1, Avenue du G\'en\'eral Leclerc, 35042 Rennes Cedex, France\\}

\date{\today}

\begin{abstract}

Sub-Kelvin scanning tunnelling spectroscopy in the Chevrel Phases {\Sn} and {\Pb} reveals two distinct superconducting gaps with $\Delta{_1}$ = 3~meV, $\Delta{_2}$~$\sim$~1.0~meV and $\Delta{_1}$ = 3.1~meV, $\Delta{_2}$~$\sim$~1.4~meV respectively.  The gap distribution is strongly anisotropic, with $\Delta{_2}$ predominantly seen when scanning across unit-cell steps on the (001) sample surface.  The spectra are well-fitted by an anisotropic two-band BCS $s$-wave gap function.  Our spectroscopic data are confirmed by electronic heat capacity measurements which also provide evidence for a twin-gap scenario.

\end{abstract}

\maketitle

Among the vast zoo of poorly-understood superconductors, Chevrel Phases (CP) stand out for their high upper critical fields $H_{c2}$, many of which exceed the Pauli limit~\cite{Fischer-1978}.  These materials were first synthesised in 1971~\cite{Chevrel-1971} and enjoyed a wealth of attention in the early 1980s.  Unfortunately, the discovery of the cuprate superconductors largely swept CP under the laboratory carpet, despite a lack of detailed understanding of their large $H_{c2}$ values.  A multi-band scenario (incorporating strong-coupling effects and enhanced spin-orbit scattering) was suggested as a possible explanation~\cite{Decroux-1982}, but until now this hypothesis has remained experimentally unexplored.  

Multi-band superconductivity was first proposed 50 years ago as a potential avenue for increasing critical temperatures~\cite{Suhl-1959}.  Interband scattering between non-degenerate bands at the Fermi level $E_F$ enables superconductivity to be induced in bands which may not directly participate in the pairing mechanism, thus increasing the effective density of states (DoS) and hence the transition temperature $T_c$.  However, with the exception of some transition metal calorimetric data~\cite{Shen-1965} and tunnelling in doped SrTiO$_3$~\cite{Binnig-1980}, multi-band superconductivity remained a largely theoretical concept until the discovery of MgB$_2$ in 2001 revived interest in the field~\cite{Nagamatsu-2001}.  In this material, superconductivity in the quasi-2D $\sigma$-band induces coherence in the quasi-3D $\pi$-band with an unexpectedly high $T_c$ of 39~K.  The two gaps have been imaged by a variety of techniques, including local spectroscopic~\cite{Eskildsen-2003} and bulk thermodynamic approaches~\cite{Bouquet-2002}.  Recently, evidence has been found for multi-band superconductivity in borocarbides~\cite{Bergk-2008}, sesquicarbides~\cite{Kuroiwa-2008}, skutterudites~\cite{Hill-2008} and, perhaps most interestingly, pnictides~\cite{Hunte-2008}.  CP and pnictides share similar anomalously large values of $H_{c2}$ and do not follow standard Werthamer-Helfand-Hohenberg theory.  However, in contrast with the pnictides, the Mo$_6$$X_8$ ($X$ = S, Se) Chevrel cluster does not exhibit any intrinsic magnetism or competing order.  This greatly simplifies the analysis and interpretation of its low-temperature properties, particularly any multi-band effects.  Bandstructure calculations have indicated the presence of two Mo $d$ bands at $E_F$ in CP~\cite{Andersen-1978}: in this Letter we present local spectroscopic evidence for two distinct superconducting gaps in {\Sn} and {\Pb}.  These data are supported by specific heat measurements displaying clear signatures of a second gap.  

We have chosen to focus on {\Sn} and {\Pb} since these two materials have the highest values for $T_c$ and $H_{c2}$ within the CP family: 14.2~K, $\sim$~40~T and 14.9~K, $>$~80~T respectively~\cite{Hc2Note}.  Single crystals of each compound with typical volume 1~mm$^3$ were grown at 1600~$^\circ$C by a chemical flux transport method using sealed molybdenum crucibles.  Their high purity was confirmed by AC susceptibility (ACS) yielding $\Delta$$T_c$~=~0.1~K for {\Sn} and 0.3~K for {\Pb}.  Local spectroscopy (STS) was performed on room-temperature-cleaved samples with a home-built helium-3 scanning tunnelling microscope in high-vacuum ($<$~10$^{-7}$~mbar), using a lock-in amplifier technique.  Heat capacity measurements were carried out at the Grenoble High Magnetic Field Laboratory with a high-resolution microcalorimeter using the ``long relaxation'' technique~\cite{Lortz-2007-2} and in Geneva using a Quantum Design$^\mathrm{TM}$ PPMS.
 
\begin{figure}[htbp]
\centering
\includegraphics [width=8.5cm,clip] {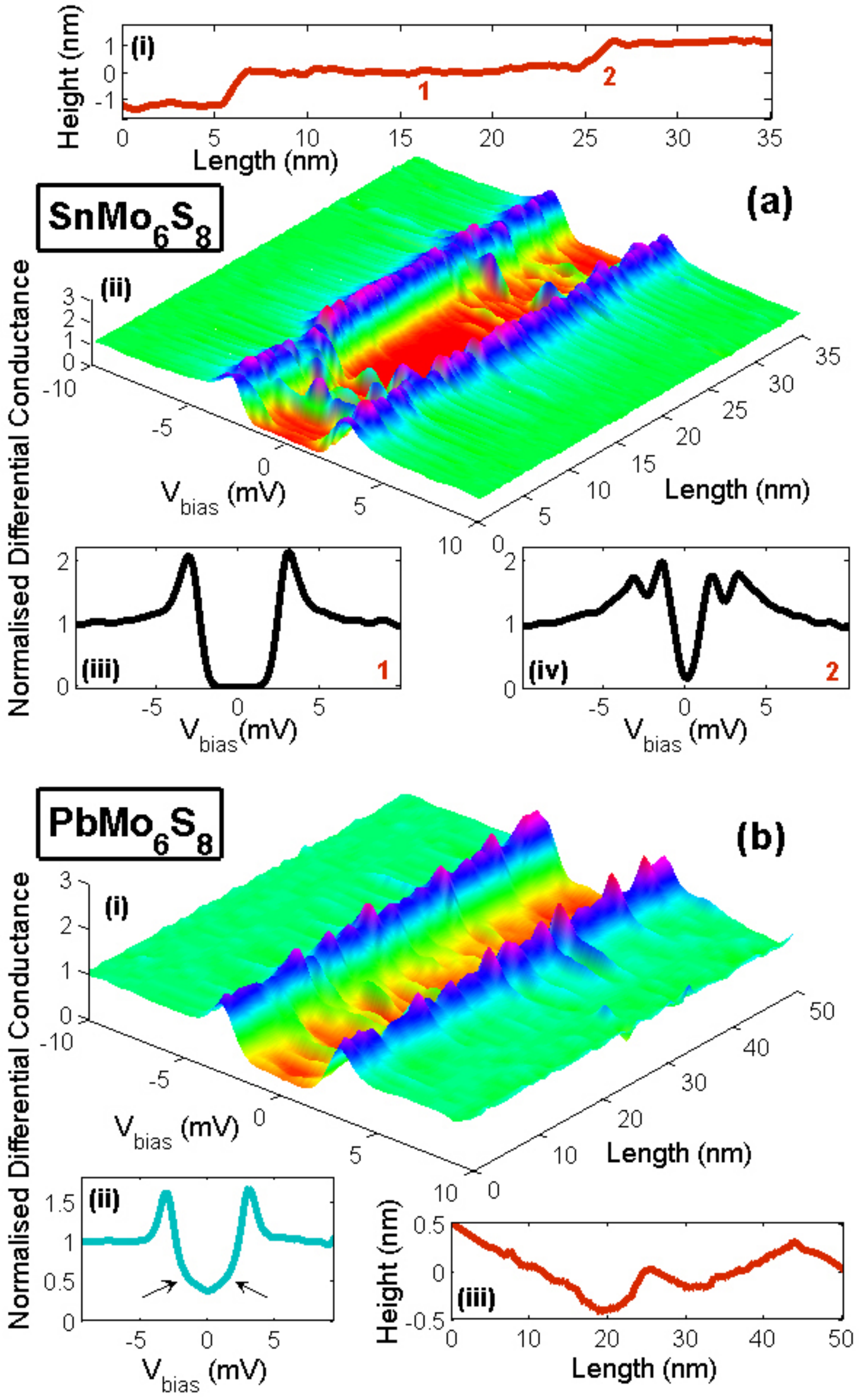}
\caption{\label{Fig_1} (a) Zero-field 35nm trace on {\Sn} taken at $T$ = 0.4K, junction resistance $R_T$ = 0.03~G$\Omega$.  (i) Topography showing double unit-cell steps; (ii) spectroscopic trace; (iii,iv) individual spectra taken on a flat terrace (1) and above a unit-cell step (2).  (b) Zero-field 40nm trace on {\Pb} taken at $T$ = 0.5K, $R_T$ = 0.015~G$\Omega$.  (i) Spectroscopic trace; (ii) average spectrum from entire trace; (iii) topographic variation.  All data are raw and unaveraged.}  
\end{figure}

The first hint of a two-band order parameter arises from fast spectroscopic traces over several tens of nanometers in the (001) plane of each material (Fig.~\ref{Fig_1}).  The corresponding topography in {\Sn} shows atomically flat terraces separated by steps of size 12~$\pm$~1~\AA, which compares favourably with twice the rhombohedral unit cell parameter 6.5~\AA.  Spectra taken on the terraces are homogeneous, with a gap of 2.95~meV and a marked lack of any quasiparticle excitations within the gap.  In contrast, spectra taken on the steps between terraces display additional peaks at lower energy, suggestive of a second gap.  We interpret this as a local modification of the tunnelling matrix element, enabling us to preferentially probe another portion of the Fermi surface with different atomic orbital characters.  Cleaved surfaces of {\Pb} are of rather lower quality with an RMS roughness of $\sim$~1.5~{\AA} and broad poorly-resolved unit cell-sized steps (Fig.~\ref{Fig_1}(b)(iii)).  However, the average spectrum (Fig.~\ref{Fig_1}(b)(ii)) displays a kink at $\sim$~$\pm$~1.4~meV (highlighted by arrows) and a V-shaped dispersion around $E_F$.  This confirms the presence of states within the large gap.  

Such a dramatic spectral variation as a function of the local topography has not previously been observed in any other superconductor.  It may therefore be natural to suggest that the isolated appearance of these multi-gap signatures at unit cell steps could be due to a surface bound state or defect.  However, a localised state would not display the particle-hole symmetry of the peaks we observe.  We have imaged a large number of separate unit cell steps and a second gap is consistently observed upon scanning across them.  Another explanation for the double-gap behaviour could be the proximity effect inducing weak superconductivity in a metallic surface layer~\cite{McMillan-1968}.  However, the small gap induced would vary strongly with the thickness of the surface metallic layer.  Apart from the fact that measurements are performed on freshly-cleaved samples, thus rendering any surface layer deposition implausible, a layer of metallic impurities would not be expected to have a uniform thickness.  This would cause substantial variation in the size of the induced gap and an extremely high zero-bias conductance (ZBC), both of which are incompatible with our data.  

\begin{figure}[htbp]
\centering
\includegraphics [width=8.5cm,clip] {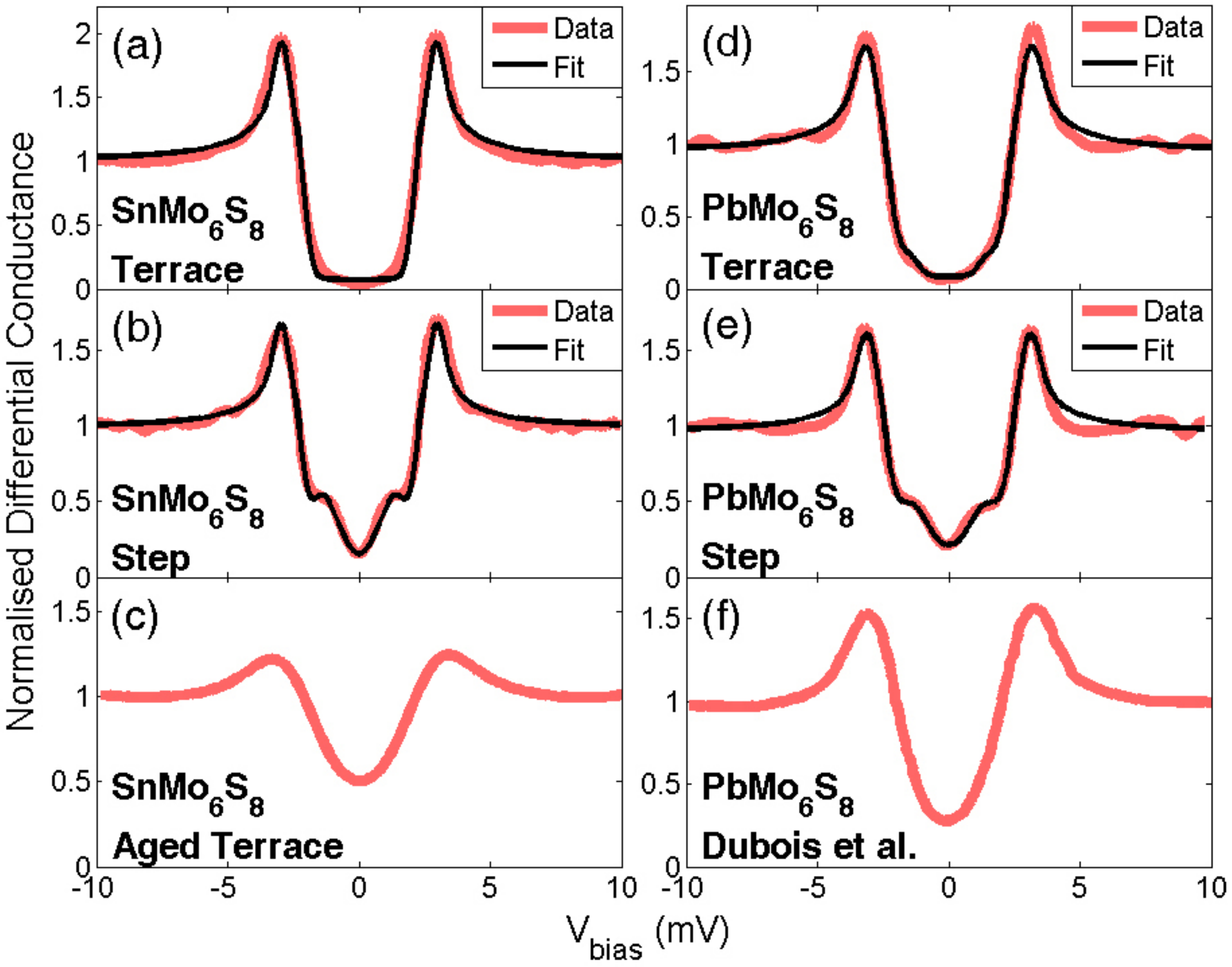}
\caption{\label{Fig_2} (a-c) {\Sn} spectra and fits taken at 0.4~K, $R_T$ = 0.03~G$\Omega$. (d-e) {\Pb} spectra  and fits taken at 0.5~K, $R_T$ = 0.015~G$\Omega$. (f) {\Pb} spectrum taken at 1.9~K, $R_T$ = 0.025~G$\Omega$ from~\cite{Dubois-2007}.  See text for details and table~\ref{Tab1} for fit parameters.}  
\end{figure}

In Figure~\ref{Fig_2} we display a range of spectra with fits using a multi-band model.  The Bardeen-Cooper-Schrieffer (BCS) quasiparticle density of states for an anisotropic $s$-wave $n$-band superconductor may be written as 
\begin{equation}\label{eq1}
N(\omega) = \sum_{j=1}^{n}\frac{N_{j}}{2\pi}\int_{0}^{\pi}\mathrm{Re}\left[\frac{(\omega+i\Gamma_{j})\mathrm{sign}(\omega)d\theta}{\sqrt{(\omega+i\Gamma_{j})^2 - \Delta_{j}^{2}F_{j}^{2}(\theta)}}\right]
\end{equation}
where $N_j$ is the contribution of band $j$ to the DoS at $E_F$, $\Gamma_j$ the scattering rate due to lifetime effects, $\Delta_{j}$ the magnitude of the gap within band $j$ and $F_{j}(\theta) = a_{j} + (1 - a_{j})\cos\theta$ measures the anisotropy of the corresponding gap with 0.5 $<$ $a_j$ $<$ 1.  We include the temperature and the experimental smearing (0.3~meV) before performing least-squares fits to our data with $N_j$, $\Delta_j$, $F_j$ and $\Gamma_j$ as free parameters.  Note that the spectral backgrounds between $\pm$~5-10~meV are rather poorly-fitted, which is indicative of strong coupling to a low-energy phonon.  However, a full Eliashberg analysis of the spectra is beyond the scope of this Letter.

Atomically flat surfaces in {\Sn} produce homogeneous spectra (Fig~\ref{Fig_2}(a)) which may be fitted using only a single band (i.e. $n$=1 in (\ref{eq1})).  There is a slight deterioration in the fit quality at low energy, which is attributed to a very small contribution from the second band.  In contrast, Fig.~\ref{Fig_2}(b) shows the average of around 50 spectra acquired above a unit cell step.  There is clearly a significant contribution from the smaller gap, necessitating a two-band fit.  Similar fits are carried out on spectra from a flat zone and a broad step in {\Pb} and the parameters obtained listed in Table~\ref{Tab1}.  We find 2$\Delta_1/k_BT_c\sim$~5 in each compound, but $\Delta_2$ is 30-40~\% larger in {\Pb} than {\Sn}.  In both materials the gap anisotropies are similar: a small anisotropy in $H_{c2}$ ($\epsilon^2$ = 0.67) has been observed in {\Pb}~\cite{Decroux-1982}, but with the present data we are unable to judge whether this is due to the anisotropy in $\Delta_1$ or $\Delta_2$.  We believe it unwise to draw quantitative conclusions on the symmetry of $\Delta_2$, since our experiment has a finite resolution imposed by a 0.3~meV broadening from the lock-in.  However, any interband scattering will preclude a pure $d$-wave order parameter in $\Delta_{2}$ due to the dominant isotropic $s$-wave component in $\Delta_{1}$.  

\squeezetable
\begin{table}[ht]
\caption{\label{Tab1} Superconducting gap parameters and relative DoS 
contributions from tunnelling (STS), heat capacity (HC) and AC 
susceptibility (ACS) data.  Both gaps $\Delta_{1,2}$ are measured in 
meV. $\Gamma_{1,2}~\leq$~0.2~meV for all STS fits.}
\begin{ruledtabular}
\begin{tabular}{lc|c|c|c|c}
{} & & \multicolumn{2}{c|}{{\Sn}} & \multicolumn{2}{c}{{\Pb}} \\
\hline
{} & technique & \multicolumn{2}{c|}{{bulk}} & \multicolumn{2}{c}
{{bulk}} \\
\hline
$T_c$ & ACS & \multicolumn{2}{c|}{14.2~$\pm$~0.05~K} & \multicolumn{2}
{c}{14.9~$\pm$~0.15~K} \\
$H_{c2}$ & HC & \multicolumn{2}{c|}{42~$\pm$~1~T} & \multicolumn{2}{c}
{86~$\pm$~5~T} \\
$\gamma$ & HC & \multicolumn{2}{c|}{6.4~$\pm$~0.1~mJgat$^{-1}$K
$^{-2}$} & \multicolumn{2}{c}{6.7~$\pm$~0.1~mJgat$^{-1}$K$^{-2}$} \\
$\Delta_1$ & HC & \multicolumn{2}{c|}{3.06~$\pm$~0.1} & \multicolumn{2}
{c}{3.15~$\pm$~0.1} \\
$\Delta_2$ & HC & \multicolumn{2}{c|}{0.86~$\pm$~0.1} & \multicolumn{2}
{c}{1.41~$\pm$~0.1} \\
$N_1$ & HC & \multicolumn{2}{c|}{96~$\pm$~2\%} & \multicolumn{2}{c}{90~
$\pm$~2\%} \\
$N_2$ & HC & \multicolumn{2}{c|}{4~$\pm$~2\%} & \multicolumn{2}{c}{10~$
\pm$~2\%} \\
\hline
{} &  & Terrace & Step & Terrace & Step \\
\hline
$\Delta_1$ & STS & 2.92~$\pm$~0.1 & 2.95~$\pm$~0.1 & 3.14~$\pm$~0.15 & 
3.06~$\pm$~0.15 \\
$\Delta_2$ & STS & -- & 1.05~$\pm$~0.2 & 1.42~$\pm$~0.2 & 1.36~$\pm
$~0.2 \\
$a_1$ & STS & 0.85~$\pm$~0.02 & 0.87~$\pm$~0.02 & 0.85~$\pm$~0.02 & 
0.89~$\pm$~0.02 \\
$a_2$ & STS & -- & 0.91~$\pm$~0.1 & 0.92~$\pm$~0.1 & 0.75~$\pm$~0.1 \\
$N_1$ & STS & -- & 62~$\pm$~4\% & 90~$\pm$~4\% & 66~$\pm$~4\% \\
$N_2$ & STS & -- & 38~$\pm$~4\% & 10~$\pm$~4\% & 34~$\pm$~4\% \\
\end{tabular}
\end{ruledtabular}
\end{table}

Previous STS experiments on {\Pb} provided evidence for low-energy excitations within the superconducting gap, but lacked sufficient resolution to distinguish two separate gaps.  This is due to three factors: sample age, temperature and environment.  In~\cite{Dubois-2007}, measurements were performed on old crystals at 1.9~K in an exchange gas, compared with freshly-grown samples at 0.4-0.5~K and high vacuum in the present work.  The increased thermal broadening at 1.9~K blurs the two gaps, though this should not be sufficient to render the smaller gap invisible.  The major factor here is a deterioration in the sample surface due to the exchange gas environment.  It is well-known that in a two-band superconductor, interband scattering due to impurities mixes the two gaps and reduces $T_c$, resulting in an effective single-band anisotropic superconductor in the dirty limit.  This was first predicted for MgB$_2$ ~\cite{Liu-2001,Mazin-2002} and later observed in irradiated samples~\cite{Wang-2003}.  However, due to extremely weak scattering between $\sigma$ and $\pi$ bands, the single band limit is never reached in MgB$_2$.  This may not be the case for CP: Fig.~\ref{Fig_2}(c) displays a {\Sn} spectrum from a terrace after 3 months of measurements comprising numerous thermal and magnetic cycles.  It is qualitatively similar to the results of~\cite{Dubois-2007} (shown in Fig.~\ref{Fig_2}(f)), providing good evidence for low-energy states within the large gap, but does not display a distinct smaller gap.  This is consistent with the presence of strong interband surface scattering.  The ZBC is also rather high in both (c) and (f), which we attribute to a decrease in the superfluid density due to enhanced pair-breaking from inelastic scattering.

\begin{figure}[b]
\centering
\includegraphics [width=8.5cm,clip] {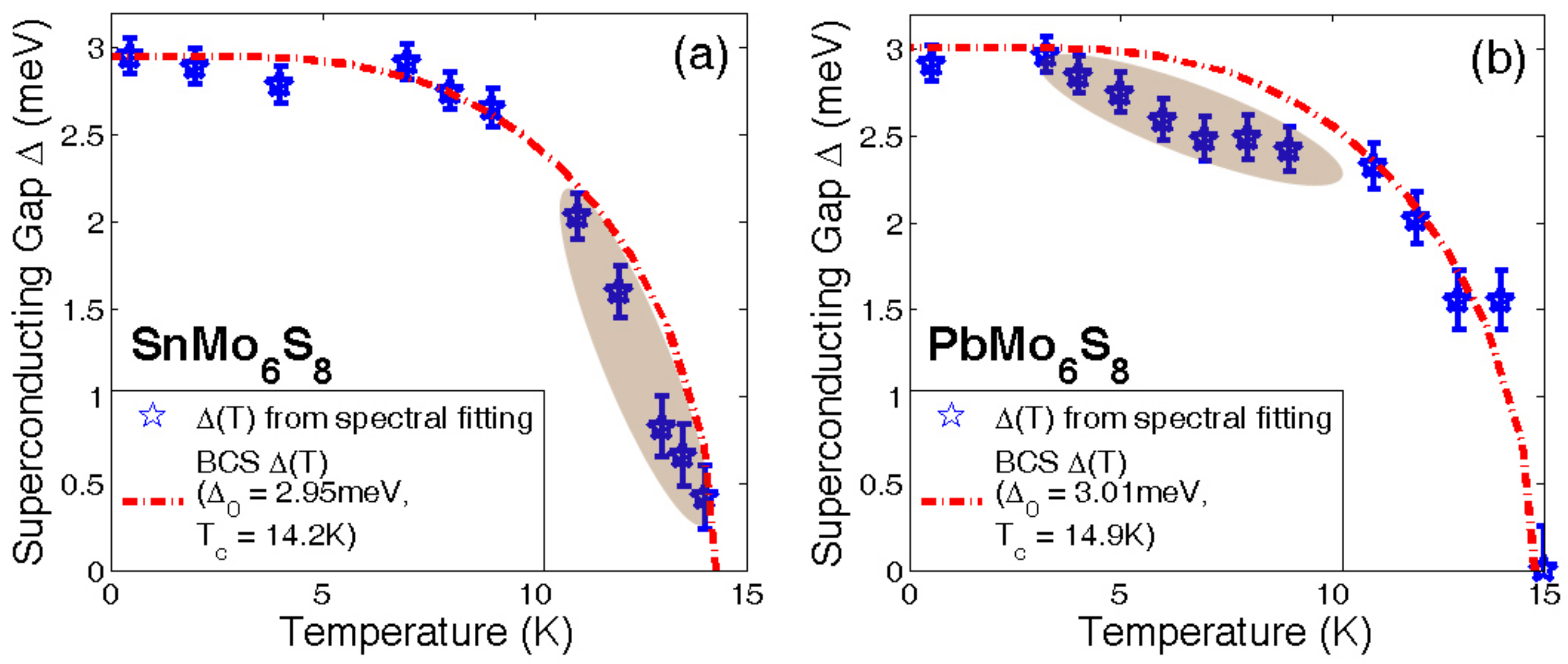}
\caption{\label{Fig_3} Variation of $\Delta_{1}(T)$ in (a) {\Sn} and (b) {\Pb}.  The gap was determined by fitting with a BCS single-band anisotropic $s$-wave model for simplicity and all spectra were acquired on a flat terrace.}  
\end{figure}

Upon increasing the temperature the large gap is gradually reduced and closes at the bulk $T_c$ determined by ACS.  No pseudogap is visible above $T_c$, confirming that superconductivity arises from a metallic ground state and hence justifying the use of a BCS model to fit the spectra.  In Fig.~\ref{Fig_3} we have plotted the variation of the large gap $\Delta_1$ with temperature for each compound, with the theoretical BCS weak-coupling $s$-wave curve for comparison.  A small kink is visible within each curve (shaded areas).  Similar features have been observed for the $\pi$-band (smaller) superconducting gap in MgB$_2$ and originate from interband scattering ``stretching'' the effective $T_c$ of the weakly-coupled $\pi$ band to the bulk $T_c$.  We suggest that in CP a small contribution from a weakly-coupled band ``stretches'' the intrinsic $T_c$ of a strongly-coupled band (which provides the majority of the DoS) to the measured bulk $T_c$.  The position of the kink at higher energy and lower temperature in {\Pb} compared with {\Sn} is consistent with our observation that band 2 is more strongly-coupled in {\Pb}.  We hypothesise that this may be the key to {\Pb} having a significantly higher $H_{c2}$ than {\Sn}, although further experiments will be required for confirmation.

\begin{figure}[htbp]
\centering
\includegraphics [width=8.5cm,clip] {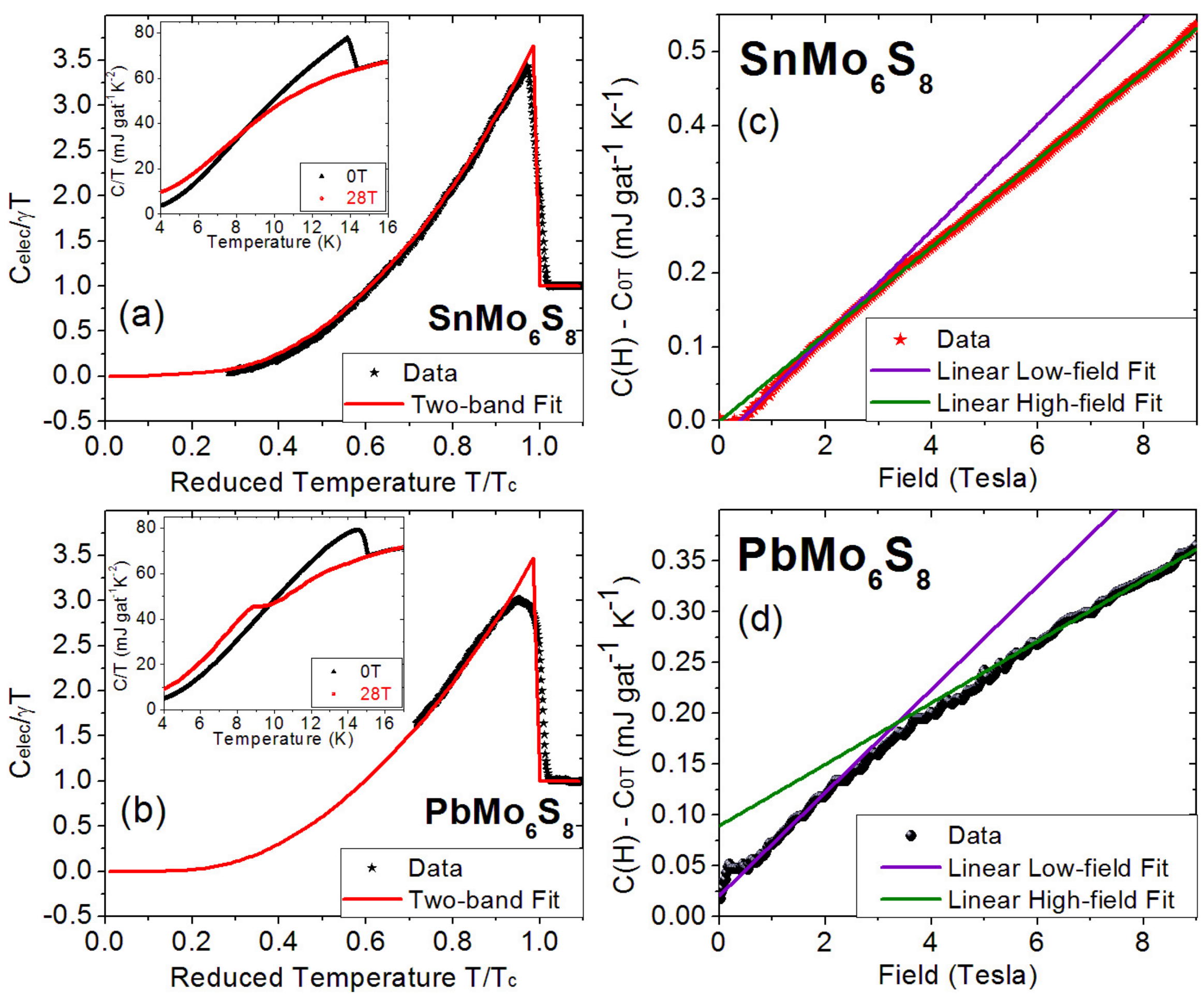}
\caption{\label{Fig_4} (a),(b) $C_\mathrm{elec}/{\gamma}T$ with two-band $\alpha$-model fits~\cite{Padamsee-1973}.  Insets: $C/T$ at 0~T and 28~T.  (c),(d) Field-dependent contributions to $C_\mathrm{elec}$ at $T$=0.35~K with linear fits above and below the crossover field $H_x$ (see text).  The crossover in {\Pb} is rather broad compared to {\Sn}; this is due to the lower homogeneity in the {\Pb} crystal as seen by increased transition widths in HC and ACS data.}
\end{figure}

It is instructive to complement our STS measurements with bulk thermodynamic (HC) data, in order to conclusively rule out any spurious surface effects being responsible for $\Delta_2$.  Figure~\ref{Fig_4}~(a) and (b) display the electronic heat capacity $C_\mathrm{elec}$ in {\Sn} and {\Pb}: this is measured by subtracting the HC in an applied field $H$~=~28~T from the zero-field data $C_\mathrm{0T}$.  To eliminate the effect of fluctuations above $T_c$ in high field, we limit our data to $T > 1.15~T_c$(28T), where $T_c$(28T) = 4.4~K and 9.3~K in {\Sn} and {\Pb} respectively.  In CP there is a large contribution to the lattice HC $C_\mathrm{latt}$ from a low-energy Einstein phonon due to vibrations of the cation between the Mo$_6$$X_8$ clusters: it is therefore not possible to calculate $C_\mathrm{latt}$ and hence extract $C_\mathrm{elec}$ using the conventional $\Sigma_{j=1}^{n}~B_{2j+1}T^{2j+1}$ model.  However, the Sommerfeld constant $\gamma$ may still be calculated using entropy considerations (see Table~\ref{Tab1}).  We find that $\gamma$/$T_c$~=~0.45~mJgat$^{-1}$K$^{-3}$ in each compound, suggesting that $T_c$ scales with the DoS at $E_F$.  Using a two-band $\alpha$-model~\cite{Padamsee-1973} we have performed fits to $C_\mathrm{elec}$ and summarise our results in Table~\ref{Tab1}.  For both gaps, the values of 2$\Delta$/$k_B$$T_c$ from our STS data agree perfectly with those from HC experiments.  While it is not possible to quantitatively compare our STS-measured $N_j$ (which also depends on the tunnelling matrix element) with the bulk $N_j$, the trends observed by each technique ($N_1$$>$$N_2$) are qualitatively in agreement.  

The final signature of 2-band superconductivity is provided by the low-temperature variation of $C_\mathrm{elec}(H)$ in each material.  In a single-band BCS $s$-wave superconductor $\gamma(H)$ should be linear.  However, at $T$~=~0.35~K we observe bends in $C_\mathrm{elec}(H)$ at $H_x$~=~2.8~$\pm$~0.2~T and 3.4~$\pm$~1~T in {\Sn} and {\Pb} respectively, reminiscent of the low-field behaviour of MgB$_2$.  Extrapolating the high-field linear fits to the normal-state value for $\gamma$, we obtain $H_{c2}$~=~42~$\pm$~1~T and 86~$\pm$~5~T (although it should be noted that these may be slight overestimates due to vortex overlap effects at high field).  We assume that $H_x$ corresponds to the crossover between filling $\Delta_2$ in band 2 followed by $\Delta_1$ in band 1 and hence estimate $N_1$=93~$\pm$~0.5~\%, $N_2$=7~$\pm$~0.5~\% and $N_1$=96~$\pm$~1~\%, $N_2$=4~$\pm$~1~\% for {\Sn} and {\Pb}.  These figures are in good agreement with those in Table~\ref{Tab1}.  

Together, our spectroscopic and thermodynamic data provide compelling evidence for a multi-band order parameter in CP superconductors.  In both {\Sn} and {\Pb}, a strongly-coupled quasi-isotropic band (contributing the majority of the DoS at $E_F$) coexists with a highly anisotropic weakly-coupled minority band.  Looking ahead, we postulate that understanding and manipulating the interplay between two or more such bands may hold the secret to realising high values for $H_{c2}$ in future superconducting materials.

\end{document}